\documentstyle[12pt]{article}
\advance\textheight by 60pt
\advance\voffset by -40pt
\advance\textwidth by 50pt
\advance\oddsidemargin by -25pt
\advance\evensidemargin by -25pt

\def\single_space{\baselineskip 12pt plus 1pt minus 1pt}
\def\one_and_a_half_space{\baselineskip 19pt plus 1pt minus 1pt}
\def\double_spacesp{\baselineskip 25pt plus 2pt minus 2pt}

\def\atversim#1#2{\lower0.7ex\vbox{\baselineskip\zatskip\lineskip\zatskip
  \lineskiplimit 0pt\ialign{$\matth#1\hfil##\hfil$\crcr#2\crcr\sim\crcr}}}

\begin{document}
\begin{titlepage}
\begin{flushright}
{\bf
PSU/TH/199\\
May 1998
}
\end{flushright}
\vskip 1.5cm
{\Large
{\bf
\begin{center}
Radiation zeros and  scalar particles \\
beyond the standard model
\end{center}
}}
\vskip 1.0cm
\begin{center}
M.~A.~Doncheski \\
Department of Physics \\
The Pennsylvania State University\\
Mont Alto, PA 17237  USA \\
\vskip 0.1cm
and \\
\vskip 0.1cm
R.~W.~Robinett \\
Department of Physics\\
The Pennsylvania State University\\
University Park, PA 16802 USA\\
\end{center}
\vskip 1.0cm
\begin{abstract}
 
Standard radiation zeros arise from the factorization properties of 
tree-level amplitudes involving a massless photon and can occur when 
all charged particles in the initial and final state have the same 
sign.  We investigate how several different processes involving new 
scalar particles beyond the standard model may exhibit radiation zeros 
and how this structure might be exploited to probe their 
electromagnetic structure.  We focus on (i) unnoticed aspects of 
angular zeros in the process 
$e^{-} + e^{-} \rightarrow \Delta^{--} + \gamma$ for doubly charged 
Higgs boson (or any bilepton) production and (ii) the process 
$\gamma + e^{-} \rightarrow q + S/V$ for scalar ($S$) or vector ($V$) 
leptoquarks (LQs). We also discuss how factorized amplitudes and 
radiation zeros may appear in the gauge boson fusion production  of 
non-conjugate leptoquark pairs via 
$\gamma + W^{\pm} \rightarrow S_i + S_j^*$ in high energy 
$e^{\pm}e^{\pm}$ reactions and how the zeros affect the production 
cross-sections for various types of scalar leptoquarks.

\end{abstract}
\end{titlepage}
\double_spacesp

\noindent
{\large {\bf 1.~Introduction}}
\vskip 0.5cm

The factorization properties \cite{halzen_zero}, \cite{bbs} of 
tree-level gauge theory amplitudes involving massless gauge quanta can 
often be used to dramatically simplify the otherwise extremely complex 
expressions for many cross-sections of physical interest.  In some 
circumstances they can also be seen to encode information on subtle 
physical effects, such as the destructive interference which 
leads to radiation zeros \cite{original} in such processes as 
$q+ \overline{q}' \rightarrow W + \gamma$, which arise from the 
interplay between several competing Feynman diagrams.  Furthermore, 
processes with the highly nontrivial angular dependences implied by 
the existence of radiation zeros are often touted as excellent sources 
of information on the electromagnetic structure of the produced 
particles, as non-gauge couplings often destroy the cancellations 
necessary for the appearance of angular zeros.  In this note we make 
use of both the factorization properties of gauge amplitudes and the 
possible existence of angular radiation zeros to discuss several 
processes involving scalar bosons beyond the standard model.  We 
discuss several such processes in which the factorized form of the 
matrix elements involved can, if nothing else, help to easily explain the 
magnitude of the total production cross-sections as one varies the 
electroweak couplings from model to model, while in several cases we see 
how the presence of angular zeros can be used to probe the electromagnetic 
couplings of such particles.

\vskip 0.5cm
\noindent
{\large {\bf 2.~$e^- + e^- \rightarrow \Delta^{--} + \gamma$}}
\vskip 0.5cm

While high-energy $e^+e^-$ or $\mu^+\mu^-$ colliders offer the 
potential for a wide variety of discoveries beyond the standard model, 
similar like-sign collisions, $e^-e^-$ or $\mu^{-}\mu^{-}$, offer a 
more specialized set of tests, such as probes of a strongly 
interacting electroweak sector \cite{han}, selectron or chargino pair 
production \cite{feng} via 
$e^{-} + e^{-} \rightarrow \tilde{e}^{-} +\tilde{e}^{-}, 
\chi^{-} + \chi^{-}$ via $t$-channel neutralino or sneutrino exchange, 
and searches for exotic Higgs bosons \cite{gunion_1}, 
\cite{alanakyan_1} or more generic doubly charged bileptons, either 
gauge bosons \cite{frampton_1} or otherwise \cite{cuypers_1}, 
\cite{raidal}. The resonant production of suitably light, doubly 
charged Higgs bosons $\Delta^{--}$ (as, for example, in various 
versions of left-right models \cite{dutta}) has been discussed 
\cite{gunion_1} and given the expected excellent energy resolution of 
muon colliders, `factory-like production' \cite{gunion_1} of such 
particles might be possible, provided the total width is small enough. 
Once discovered in this way, more detailed studies of processes such 
as $e^- + e^- \rightarrow \Delta^{--} + \gamma$ could serve to probe 
the electromagnetic properties of such a state (or any doubly charged 
bilepton). Alanakyan \cite{alanakyan_2} has recently calculated the 
cross-section for this process, correcting a much earlier prediction 
\cite{rizzo} by taking all tree-level diagrams into account, and 
obtains the expression 
\begin{equation}
\frac{d\sigma}{d\cos(\theta)} 
= \frac{\alpha h_{ee}^2}{s}\left(1 + \frac{2(1-\beta)}{\beta^2}\right)
\beta \cot^2(\theta)
\end{equation}
where $\beta = 1 - M_{\Delta}^2/s$ and $h_{ee}$ is the appropriate 
$ee\Delta^{--}$ coupling. The author of Ref.~\cite{alanakyan_2} 
focuses attention on the collinear singularities (when 
$\theta = 0,\pi$) in the cross-section, which are reminiscent of those 
in $e^{+} + e^{-} \rightarrow Z^{0} + \gamma$, but does not mention 
the other dramatic angular dependence, namely the angular zero in the 
tree-level cross-section when $\theta = \pi/2$.  Given the general 
arguments of Refs.~\cite{halzen_zero} and \cite{bbs}, we would expect 
the cross-section for this process to factorize with a pre-factor 
containing all of the relevant charge factors.  Repeating the 
calculation for this process with this in mind, we find an entirely 
equivalent expression given by 
\begin{equation}
\frac{d \sigma}{d t}
= \frac{\alpha h_{ee}^2}{s^2}
[(Q_{1}u - Q_{2}t)^2]
\left[\frac{s^2 + M_{\Delta}^4}{ut(u+t)^2}\right]
\label{ee_zero}
\end{equation}
where $t = -s\beta(1-\cos(\theta))/2$ and we have chosen to write 
$Q_1 = Q_2 \equiv Q_e$ and to separate the pre-factor in square 
brackets including the initial state charges to emphasize its origin 
in the factorization of gauge theory amplitudes involving massless 
photons.  The resulting characteristic angular zero at $90^{\circ}$ in 
the center-of-mass frame can obviously be used to probe the 
electromagnetic structure of the $\Delta^{--}$ and we will briefly 
discuss this further below. 

This angular zero has already been noticed,  in the context of doubly 
charged vector bileptons, $V^{--}$, by Cuypers and Raidal 
\cite{cuypers_1} who note that its presence depends the standard gauge 
coupling of the $V^{--}$ and will be `filled in' if the bilepton has 
an anomalous magnetic moment coupling given by $\kappa \neq 1$.  To 
focus on this in more depth, we note that the cross-section for 
$e^{-} + e^{-} \rightarrow V^{--} + \gamma$ for a general value of 
such an anomalous coupling is proportional to
\begin{eqnarray}
\frac{d \sigma}{dt}
& \propto &
[(Q_{1} u - Q_{2}t)^2]\left[\frac{u^2 + t^2 + 2s M_V^2}{ut(u+t)^2}\right]
+ (\kappa -1) [Q_{1} u - Q_{2} t]\left[\frac{t-u}{(t+u)^2}\right] 
\nonumber \\
& & 
\qquad
\qquad
\qquad 
+ \frac{(\kappa-1)^2}{2(t+u)^2}
\left[tu + (t^2+u^2)\frac{s}{4M_V^2}\right]
\label{vector_zero}
\end{eqnarray}
where we once again keep separate those terms proportional to 
$(Q_{1} u - Q_{2}t)$ arising from factorization and terms of the form 
$(t-u)$ which arise from other kinematical effects.  We note that in 
the very special case of $e^{-}e^{-},\mu^{-}\mu^{-}$ collisions where 
$Q_{1} = Q_{2}$, not only will the standard gauge theory cross-section 
term vanish at the radiation zero as $(Q_1u-Q_2t)^2 = (u-t)^2$, {\it i.e.} 
as two powers of $\cos(\theta)$, but that the interference term will do so as 
well, in contrast the most general case where the cross-term will only 
be suppressed by one power of the $(Q_1u - Q_2t)$ pre-factor. This 
implies that the cross-section near the angular location of the 
radiation zero is additionally sensitive to new physics contained in 
the $(\kappa-1)^2$ term in the special case of equal initial charges.

Returning to doubly charged Higgs production, we note that if we 
parameterize the $\Delta^{--}$ electromagnetic coupling with a 
form-factor given by $F(q^2) \approx 1- \delta(q^2)$, the angular 
distribution in Eqn.~(\ref{ee_zero}) will now contain terms of order 
$\delta$ and $\delta^2$. The interference term proportional to 
$\delta$ turns out to have a kinematic factor which is also 
proportional to $(t-u)$, just as for the vector case, leaving the 
radiation zero in $e^- + e^- \rightarrow \Delta^{--} + \gamma$ more 
sensitive to the presence of non-pointlike electromagnetic couplings 
through the remaining $\delta^2$ term which does not vanish.

\vskip 0.5cm
\noindent
{\large {\bf 3.~$\gamma + e^{-} \rightarrow q + S$}}
\vskip 0.5cm

Resonant leptoquark production via the process $e + q \rightarrow S/V$ 
for both scalar ($S$) and vector ($V$) leptoquarks has been considered 
in the context of $ep$ collisions at HERA and beyond any number of 
times. A logical extension to the process 
$e + q \rightarrow S/V + \gamma$ has been analyzed in some detail 
\cite{raz_ep} and the presence of the radiation zero and its 
dependence on the electromagnetic couplings of the leptoquark have 
been discussed.  (We note that various studies of radiation zeros in 
non-resonant $e+q \rightarrow e+q+\gamma$ scatterings have also 
appeared \cite{non_resonant}.)  The radiative decays of scalar 
leptoquarks \cite{deshpande}, via a crossed version of this process, 
namely $S \rightarrow e + q + \gamma$, have also been considered as an 
extension of processes involving charged scalar particles 
\cite{scalars} such as $q\overline{q}' \rightarrow H^{\pm} \gamma$, 
$H^{\pm} \rightarrow q\overline{q}' \gamma$, and 
$H^{\pm} \rightarrow \tilde{q} (\tilde{q'})^* \gamma$ (where 
$\tilde{q}$ are scalar quarks.)

The other crossing of the basic process, namely 
$\gamma + e \rightarrow S + q$, is of relevance to single scalar 
leptoquark production in $ee$ and $e\gamma$ collisions and has been 
discussed by Hewett and Pakvasa \cite{hewett} (for the special case of 
$Q_S = -1/3$) and then generalized by Nadeau and London \cite{nadeau} 
for arbitrary $Q_S$. Neither group presents their results in a 
factorized form of the type guaranteed by the arguments of 
Ref.~\cite{halzen_zero} and the general cross-section is presented in 
Ref.~\cite{nadeau} as 
\begin{equation}
\frac{d \sigma}{d \hat{u}}
= -\frac{\pi k \alpha^2_{em}}{2 \hat{s}^2}F(\hat{s},\hat{t},\hat{u},M_S^2)
\end{equation}
where
\begin{eqnarray}
F(\hat{s},\hat{t},\hat{u},M_S^2) & = &
- \frac{\hat{u}}{\hat{s}} 
- 2Q_S^2 \frac{\hat{u}(\hat{u} + \hat{s} - M_S^2)^2}
{\hat{s}(\hat{u}+\hat{s})^2} 
- 2Q_S \frac{\hat{u} (\hat{u}+\hat{s} -M_S^2)}{\hat{s}(\hat{u}+\hat{s})}
+ 2(1+Q_S) \frac{(\hat{u} - M_S^2)}{\hat{s}}
\nonumber 
\\
& &
- (1+Q_S)^2 
\left[\frac{\hat{s}}{\hat{u}} + 2\frac{(\hat{u} - M_S^2)(\hat{u}+\hat{s} - M_S^2)}{\hat{u}\hat{s}} \right] 
\label{f-function}
\\
& & 
+ 4Q_S(1+Q_S) \frac{(\hat{u}+\hat{s} - M_S^2)
(\hat{s}/2 + \hat{u} - M_S^2)}{\hat{s}(\hat{u}+ \hat{s})} \nonumber 
\end{eqnarray}
In this expression, $\hat{t} = -\hat{s}\beta(1-\cos(\theta^*))/2$, 
$\beta = 1-M_S^2/\sqrt{s}$, and $\theta^*$ is the angle between the 
produced $S$ ($q$) and the incident $\gamma$ ($e^{-}$) in the 
center-of-mass frame; the $l\!-\!q\!-\!S$ coupling is given as $g$ and 
one defines $g^2/4\pi \equiv k \alpha_{em}$.

Using the arguments in Ref.~\cite{halzen_zero} or \cite{bbs} we know 
that this expression can be factorized and either by direct 
calculation or algebraic manipulation of Eqn.~(\ref{f-function}) we 
indeed find that the differential cross-section can be written in a 
very simple form, namely
\begin{equation}
\frac{d \sigma}{d \hat{u}}
= \frac{\pi k \alpha_{em}^2}{2 \hat{s}^2}
\left[(1+Q_S)\hat{s} + \hat{u}\right]^2 
\left[
\frac{\hat{t}^2 + M_S^4}{\hat{s}\hat{u}(\hat{u} + \hat{s})^2} 
\right]
\label{simple_form}
\end{equation}
(Note the connection to the cross-section in Eqn.~(\ref{ee_zero}) 
which is basically the crossed process, but with different values of 
the charges of the particles.)  The factorization of the entire $Q_S$ 
dependence into the term in square brackets helps further explain the 
relative magnitude of the integrated cross-sections (after appropriate 
$p_T$ cuts) seen in Ref.~\cite{nadeau} for various values of $Q_S$.  
There the authors note that the cross-sections for the two processes 
involving $|Q_q| = 2/3$ quarks (i.e., those with $Q_S=-5/3,-1/3$) are 
larger  than for those involving $|Q_q| = 1/3$ quarks (corresponding 
to $Q_S=-4/3,-2/3$) due to the resulting enhanced coupling of photons 
to the quarks in the $t$-channel exchange diagram.  Using the 
factorization above, we can also see that the $Q_S = -5/3$ and 
$Q_S=-1/3$ cross-sections themselves will differ only in the 
pre-factors given by $[-2\hat{s}/3+\hat{u}]^2$ and 
$[+2\hat{s}/3 + \hat{u}]^2$ and the first term is larger since 
$\hat{u} < 0$; a similar hierarchy is then present for the 
$Q_S = -4/3$ and $Q_S=-2/3$ cases.

While the authors of Refs.~\cite{hewett} and \cite{nadeau} present 
expressions for total cross sections (again, after appropriate cuts), 
they do not discuss the angular zeros which may be present in these 
processes.  (We note that Cuypers \cite{cuypers_rad_zero} has 
discussed possible radiation zeros in this process, but does not 
report the ``long analytical forms'' for the cross-sections which we 
have found here to factorize very simply.)  We note that the 
pre-factor in square brackets can vanish when
\begin{equation}
\cos(\theta^*) \equiv y = \frac{2(1+Q_S)}{\beta} -1
\qquad
\longrightarrow
\qquad
1+ 2Q_S
\qquad
\quad 
\mbox{for $\beta \rightarrow 1$}
\label{physical_zero}
\end{equation}
so that there can be angular zeros when $Q_S = -1/3,-2/3$, 
corresponding to $Q_q = -2/3,-1/3$, since then all charged particles 
in the initial and final states are of the same sign.  In addition, 
zeros will only be present in the physical region provided that 
$\beta$ is large enough. The condition that a zero will appear in the 
observable range of interest, namely $-1 \leq y \leq +1$, is given by 
\begin{equation}
\frac{\sqrt{\hat{s}}}{M_S} \equiv z \geq \frac{1}{\sqrt{-Q_S}}
\label{min_beta}
\end{equation}
or $\sqrt{\hat{s}}/M_S \geq 1.73\,(1.25)$ for $Q_S = -1/3\,(-2/3)$.

Since the angular dependence of the cross-section is determined by the 
dimensionless term $F(\hat{s},\hat{t},\hat{u},M_S^2)$, we plot this 
function for four values of $Q_S$ as a function of 
$y = \cos(\theta^*)$ in Fig.~1 for several different values of 
$z \equiv \sqrt{\hat{s}}/M_S$.  The differential cross-sections for 
$y\rightarrow -1$ are larger for cases (a)/(b) compared to (c)/(d) due 
to the charge of the exchanged ($t$-channel) quark ($|Q_q| = 2/3$ 
compared to $|Q_q| = 1/3$) as mentioned above, while radiation zeros 
are present in cases (b) and (d) for large enough values of $\beta$ as 
in Eqn.~(\ref{min_beta}); the angular locations of these zeros are 
seen to be consistent with Eqn.~(\ref{physical_zero}). 

Whether the photons in the $\gamma e$ collisions are ``effective'' 
(arising from an approximately real, Weisz\"acker-Williams photon in 
$e^+e^-$ colliders) or ``real'' (arising, for example, from laser 
backscattering), the angular dependence will not be probed at fixed 
center-of-mass energy.  Either mechanism provides a photon beam with a 
distribution of energies, and so a range of $\hat{z} = \sqrt{\hat{s}}/M$ 
will be probed.  Rather than a zero in certain cross sections as seen at 
the parton level (illustrated in Fig.~1), a broad (in $\cos \theta$) 
region of reduced cross section will be seen.  In Fig.~2, we show the 
results of a calculation of the lab frame angular distribution at a 
$\sqrt{s} = 1\,TeV$ $e^+e^-$ linear collider, utilizing laser 
backscattering to 
operate in $e\gamma$ mode.  The radiation amplitude zero is, indeed, 
filled in, but its effect can be seen in the shape of the 
$\cos \theta$ distribution, much as was noted in 
Refs.~\cite{non_resonant} and \cite{cuypers_rad_zero}.

The presence of angular zeros is not unique to scalar states in such 
single leptoquark production processes.  Montalvo and \`Eboli 
\cite{eboli} have considered the production of composite vector 
leptoquark states via the same mechanism, namely 
$\gamma + e^{-} \rightarrow q + V$.  They consider the interaction 
given by 
\begin{equation}
{\cal L} = -g V_{\mu}^{ab\,\dagger} \overline{L}^a \gamma^{\mu}
L^{b} + H.c.
\end{equation}
where $L^a$ are the physical $SU(2)$ left-handed doublets of the 
standard model and $Q_S = -2/3$ in the specific model considered while 
we have $g^2/4\pi = k \alpha_{em}$ as before to be consistent with the 
notation used here. If we ignore the final state quark mass, one can 
see that their result for the differential cross-section can be 
written in a form entirely analogous to Eqn.~(\ref{simple_form}), 
namely
\begin{equation}
\frac{d \sigma}{d \hat{t}}
=
- \frac{12 \pi k \alpha^2_{em}}{\hat{s}^2}
\left[(1+Q_S)\hat{s} + \hat{t}\right]^2
\left[\frac{\hat{s}^2 + \hat{t}^2 + 2\hat{u}M_V^2}
{\hat{s}\hat{t} (\hat{s} + \hat{t})^2} \right]
\label{vector_result}
\end{equation}
where the role of $\hat{t}$ and $\hat{u}$ are interchanged due to a 
differing definition of $\theta^*$ in Ref.~\cite{eboli} compared to 
the one used here.   We note that they have chosen a value of the 
coupling $g$ which, along with the standard $VV\gamma$ vertex gives 
rise to large energy behavior for $e^+e^- \rightarrow VV^*$ for which 
unitarity is maintained at tree-level.  The fact that their result in 
Eqn.~(\ref{vector_result}) is then entirely similar to the crossed 
version of the standard result for 
$q + \overline{q}' \rightarrow W + \gamma$ \cite{original} or that 
seen in Eqn.~(\ref{vector_zero}) (only differing in the factorized 
term containing the charge dependence) is then easily understood and 
angular zeros will also be present in the explicit composite model 
case they consider, subject to the kinematic condition in 
Eqn.~(\ref{min_beta}).

\vskip 0.5cm
\noindent
{\large {\bf 4.~$\gamma + W^{\pm} \rightarrow S_{i} + S_{j}^*$}}
\vskip 0.5cm

While single production of leptoquarks in $eq$ or $\gamma e$ 
collisions may well be important, the current best limits on 
leptoquark masses arise from processes involving pair production.  
Analyses from hadron colliders using $gg$ and $q\overline{q}$ fusion 
processes \cite{hewett_gg}, including appropriate NLO corrections 
\cite{nlo_leptoquark}, now routinely set limits of order 
$M(LQ) > 200\,GeV$ for leptoquarks \cite{lq_limits} with branching 
ratios to charged leptons of $BR(LQ\rightarrow eq) > 1/2$.  Similarly, 
production prospects from $e^{+}e^{-}$ \cite{leptoquark_pair_ee} and 
$\gamma \gamma$ \cite{gamma_gamma} collisions have been examined in 
great detail.  Such processes are important for the extraction of 
unambiguous mass limits as the production cross-sections depend on the 
well-defined gauge quantum numbers of the leptoquarks and not on their 
unknown couplings to $lq$ pairs.  (Such couplings can, of course, 
contribute to these pair production processes, such as from the 
$t$-channel quark exchange diagram in $e^{+}e^{-}$ collisions.) In all 
such cases, one obviously produces pairs of opposite sign, conjugate 
leptoquarks ($SS^*$ or $VV^*$) and the production cross-sections are 
essentially independent of the leptoquark generation.

Cuypers, Frampton, and R\"uckl \cite{non_conjugate} have noted that it 
is possible produce pairs of non-conjugate leptoquarks in $e^{-}e^{-}$ 
collisions via $t$-channel quark exchange under very special 
circumstances, requiring the simultaneous 
existence of both $|F| = 2$ and $F=0$ 
leptoquarks which couple with the appropriate chirality to 
first-generation leptons.  While this is an intriguing possibility, it 
requires an array of leptoquarks which only appear in very specialized 
models and relies explicitly on the unknown $LQ-l-q$ couplings which may 
very well be small, especially for the first generation.

A more standard source of production of two non-conjugate leptoquark 
pairs in either $e^{+}e^{-}$ or $e^{-}e^{-}$ collisions arises  from 
the subprocess $\gamma + W \rightarrow S_i + S^*_j$, which is possible 
provided the leptoquark (of any generation) transforms non-trivially 
under $SU(2)$.  Since many of the standard leptoquark assignments 
\cite{wyler}, \cite{much_ado} allowed by 
$SU(3)\times SU(2)\times U(1)$ invariance transform as either doublets 
or triplets, such processes will be accessible in  a much wider 
variety of models.  (In this same context, the analogous 
$\gamma + W^{+} \rightarrow t + \overline{b}$ process \cite{kauffman} 
has been considered in detail by Kauffman.)
In what follows we simply 
characterize the basic cross-section for this process, indicating how 
the amplitude factorization can simplify the resulting matrix elements 
and how the presence of radiation zeros leads to suppression of the 
cross-section for certain leptoquarks. (A complete discussion of the
pair production of leptoquarks via all gauge boson fusion processes will
appear elsewhere \cite{mike_and_rick_new}.)

We characterize the basic tree-level process as 
$\gamma(k) + W^{-}(p) \rightarrow S_i(q_1) + S^{*}_{j}(q_2)$ labeling 
the momenta.  The matrix element can be written from the beginning in 
a factorized form, namely 
\begin{equation}
{\cal M} = 4ieG\left(Q_i - Q\frac{k\cdot q_1}{k \cdot p}\right)
\left(
\frac{q_{1\mu}q_{2\nu}}{l_1^2 - m^2}
+
\frac{q_{2\mu}q_{1\nu}}{l_2^2 - m^2}
+ \frac{g_{\mu \nu}}{2}
\right) \epsilon_{\gamma}^{\mu}(k) \epsilon_{W}^{\nu}(p)
\end{equation}
where $l_1^2 = (p-q_2)^2 = (k-q_1)^2$ and 
$l_2^2 = (p-q_1)^2 = (k-q_2)^2$, $Q_i = Q(S_i) = Q(S_j) -1$, and we 
assume that the masses of the non-conjugate leptoquarks are 
degenerate, {\it i.e.} $M(S_i) = M(S_j) = m$. (Constraints from precision 
electroweak data, for example the $\rho$ parameter, limit the mass 
splittings of leptoquarks \cite{keith_ma}, especially for $SU(2)$ 
triplets \cite{mike_rick_third}.)  The overall electroweak coupling 
factor is given by $G = g/\sqrt{2},g$ for $SU(2)$ doublets and 
triplets respectively. We note that this expression confirms and 
extends an earlier expression \cite{grose} for the amplitude 
describing the radiative decay of $W$ bosons into massless scalar 
quark pairs.)

The cross-section can then be written in the form 
\begin{equation}
\frac{d\sigma}{dt}(\gamma W^{-} \rightarrow S_iS_j^*)
= \frac{8\pi \alpha^2f_W}{s^2 \sin^2(\theta_W)}
\left(Q_i + \frac{\tilde{u}}{\tilde{u} + \tilde{t}}\right)^2
G(s,\tilde{t},\tilde{u},M_W^2,m^2)
\end{equation}
where $f_W = 1/2,1$ for $SU(2)$ doublets and triplets respectively and 
\begin{eqnarray}
G(s,\tilde{t},\tilde{u},M_W^2,m^2)
& = & 
\frac{1}{4\tilde{u}^2 \tilde{t}^2}
\left[
\tilde{u} \tilde{t}[2\tilde{u}\tilde{t} + sM_W^2]
- m^2[(s-M_W^2)^2M_W^2 + 4\tilde{u}\tilde{t}s] \right. \nonumber \\
& &
\qquad \qquad \qquad
\left. + 4m^4(s-M_W^2)^2\right]
\end{eqnarray}
with $\tilde{t} = t - m^2$ and $\tilde{u} = u - m^2$.  In this simple 
form, the electroweak and electromagnetic couplings of different 
leptoquarks are easily separated into the $f_W$ and pre-factor 
containing the LQ charge. In contrast to the $\gamma \gamma$ 
cross-section for production of conjugate pairs, where there is an 
overall $Q_i^4$ factor which trivially determines the relative 
importance of the process for various leptoquark assignments, the 
interplay between the photon coupling to different diagrams here is 
slightly more complex, but still encoded in a fairly simple 
pre-factor. To see what effect varying these parameters has on the LQ 
production cross-section, we plot, in Fig.~3, the expression 
\begin{equation}
f_W \left(Q_i + \frac{1+y}{2}\right)^2
\label{angular_limit}
\end{equation}
which gives the appropriate combination of these pre-factors in the 
high energy limit for several types of scalar leptoquarks.  We plot 
this function, which determines the different dependences on charge 
and electroweak coupling in the angular distributions in the 
center-of-mass frame for four cases which can appear in $\gamma W^{-}$ 
collisions:
\begin{equation}
\begin{array}{ccccc}
\mbox{LQ type} & Q(S_i) & Q(S_j^*) & SU(2) \,\mbox{rep} & f_W \\ 
               &        &          &                        \\
R_{2L},R_{2R}  & +2/3   & -5/3     & \mbox{doublet}   & 1/2 \\
\tilde{R}_{2L} & -1/3   & -2/3     & \mbox{doublet}   & 1/2 \\
S_{3L}         & -2/3   & -1/3     & \mbox{triplet}   & 1 \\
S_{3L}         & +1/3   & -4/3     & \mbox{triplet}   & 1 \\
\end{array}
\label{lq_charges}
\end{equation}
where we use the leptoquark labeling scheme of Ref.~\cite{much_ado}.  The total 
center-of-mass cross-section can be written in the form
\begin{equation}
\sigma(s) = \frac{1}{s}
\left[\frac{4\pi \alpha^2}{\sin^2(\theta_W^2)}\right]
R(z)
\end{equation}
where
\begin{equation}
R(z) = \int_{-1}^{+1} \beta \left[f_W 
\left(Q_i + \frac{\tilde{u}}{\tilde{u} + \tilde{t}}\right)^2\right]
G(s,\tilde{t},\tilde{u},0,m^2)\, dy
\label{integrated_cross_section}
\end{equation}
where $\beta \equiv  \sqrt{1-4m^2/s}$ and 
$\tilde{t} = -s(1-\beta y)/2$, $\tilde{u} = -s(1+\beta y )/2$, 
$z \equiv \sqrt{s}/2m$.  (Note that given the existing mass limits on 
leptoquarks, we will ignore $M_W$ as we are interested in the limit 
where $s \geq 2m >> M_W$.)  We plot $R(z)$ versus $z$ in Fig.~4 for 
the leptoquark charge assignments in Eqn.~(\ref{lq_charges}) and 
Fig.~3 and we see that the differences in total cross-section are 
easily explained by the differences in pre-factors. Similar results 
for vector leptoquarks (including anomalous couplings) are easily 
obtained as generalizations and we are guaranteed that the same 
electroweak and charge pre-factors will appear in those processes as 
well, at least as long as one is restricted to purely gauge couplings.

\vskip 0.5cm
\noindent
{\large {\bf 5.~Acknowledgments}}
\vskip 0.5cm
One of us (M.A.D) acknowledges the support of Penn State University 
through a Research Development Grant (RDG).

\newpage
{\Large
{\bf Figure Captions}}
\begin{itemize}
\item[Fig.\thinspace 1.] Plots of $F(s,t,u,M_S^2)$ (defined in 
Eqn.~(\ref{f-function})) versus $y = \cos(\theta^*)$ where $\theta^*$ 
is the angle between the produced $S$ ($q$) and the incident $\gamma$ 
($e^-$) in the center-of-mass frame.  Plots for four possible 
leptoquark charges are shown and curves for center-of-mass energies 
described by $z = \sqrt{\hat{s}}/M = 5.0,2.0,1.5,1.3,1.1$ are given by the 
solid, dashed, dot-dash, dotted, and dot-dash-dash lines respectively. 
Radiation zeros are present only in the case where the charged 
particles in the final state state (the $S$ and $q/\overline{q}$) have 
the same charges as those in the initial state (the $e^{-}$), 
consistent with the general theorems of Ref.~\cite{bbs} and are in the 
physical region for sufficiently large values of $z = \sqrt{\hat{s}}/M_S$ 
(Eqn.~\ref{min_beta}) and their locations are given by 
Eqn.~(\ref{physical_zero}).
\item[Fig.\thinspace 2.] Plots of $d\sigma/d \cos \theta$, 
Eqn.~(\ref{simple_form}) convoluted with a laser backscattered photon 
distribution, versus $y = \cos(\theta)$ where $\theta$ is the angle 
between the produced $S$ ($q$) and the incident $\gamma$ ($e^-$) in 
the lab frame.  Plots for four possible leptoquark charges are shown 
and curves for leptoquark masses described by 
$z = \sqrt{s}/M = 5.0,2.0,1.5,1.3$ are given by the solid, dashed, 
dot-dash and dotted lines respectively, for leptoquark production at a 
$\sqrt{s} = 1\,TeV$ $e^+e^-$ collider operating in $e\gamma$ mode.  The 
values of $z$ were chosen for easy comparison with Fig.~1; for 
$z = \sqrt{s}/M = 1.1$, the mass of the leptoquark is very near the 
kinematic limit of the collider in $e\gamma$ mode (laser backscattering 
produces a photon beam with maximum energy slightly lower than the 
initial electron beam energy), and the cross section is tiny.
\item[Fig.\thinspace 3.] Plot of Eqn.~(\ref{angular_limit}) which 
describes the dependence of non-conjugate leptoquark production on the 
electroweak and electromagnetic couplings in the high energy limit, 
namely $F_W(Q_i+(1+y)/2)^2$ versus $y$.  The cases we consider are 
denoted by the values of $(Q_i,f_W)$ given by $(+1/3,1)$ (solid), 
$(+2/3,1/2)$ (dashed), $(-2/3,1)$ (dot-dash), and $(-1/3,1/2)$ 
(dotted).
\item[Fig.\thinspace 4.] Plot of $R(z)$ (which gives the integrated 
cross-section via Eqn.~(\ref{integrated_cross_section})) versus 
$z = \sqrt{s}/2m$ for four different cases of non-conjugate leptoquark 
production.  The cases considered are the same as those in Fig.~3, 
namely $(+1/3,1)$ (solid), $(+2/3,1/2)$ (dashed), $(-2/3,1)$ 
(dot-dash), and $(-1/3,1/2)$ (dotted).
\end{itemize}
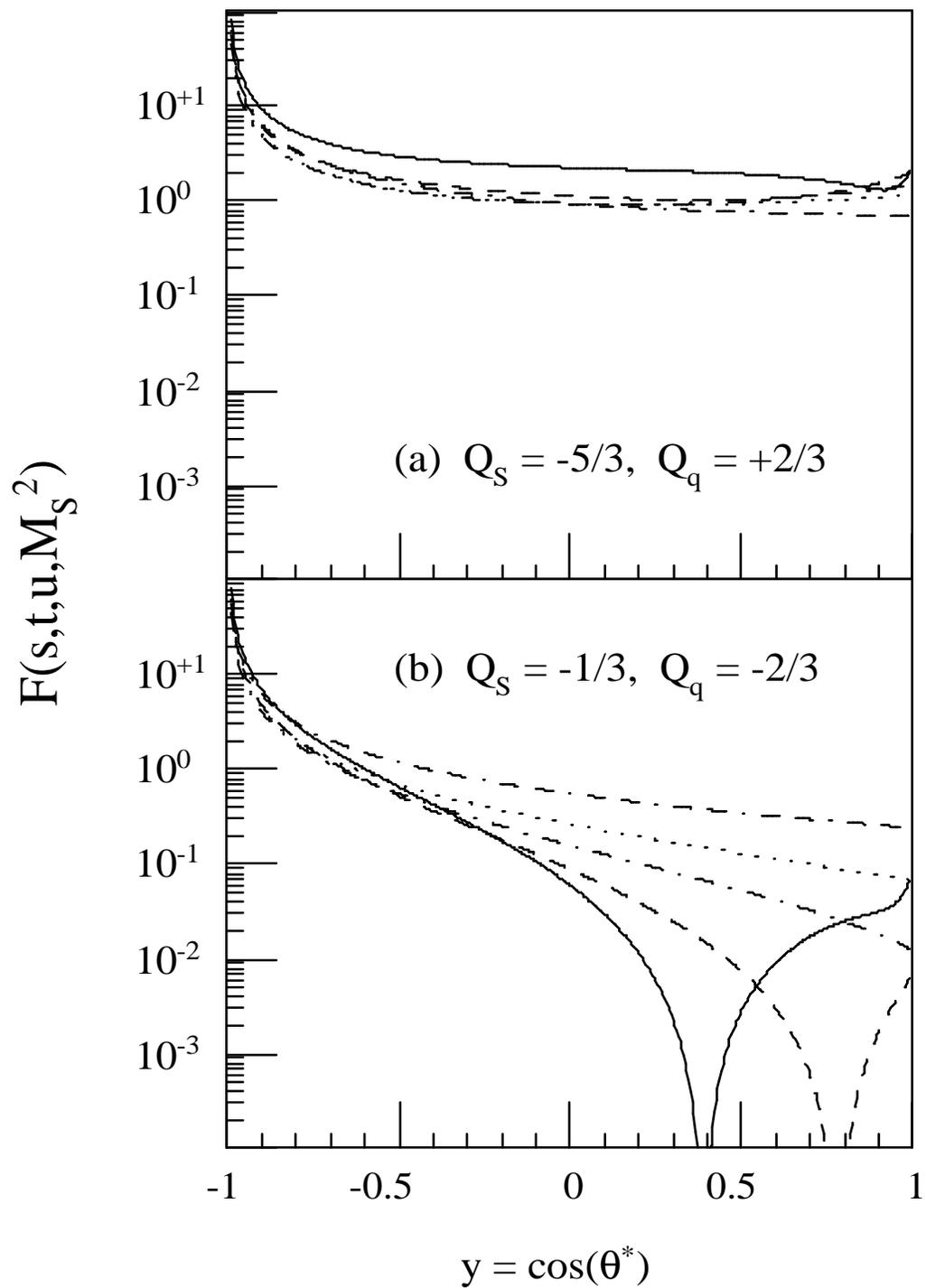
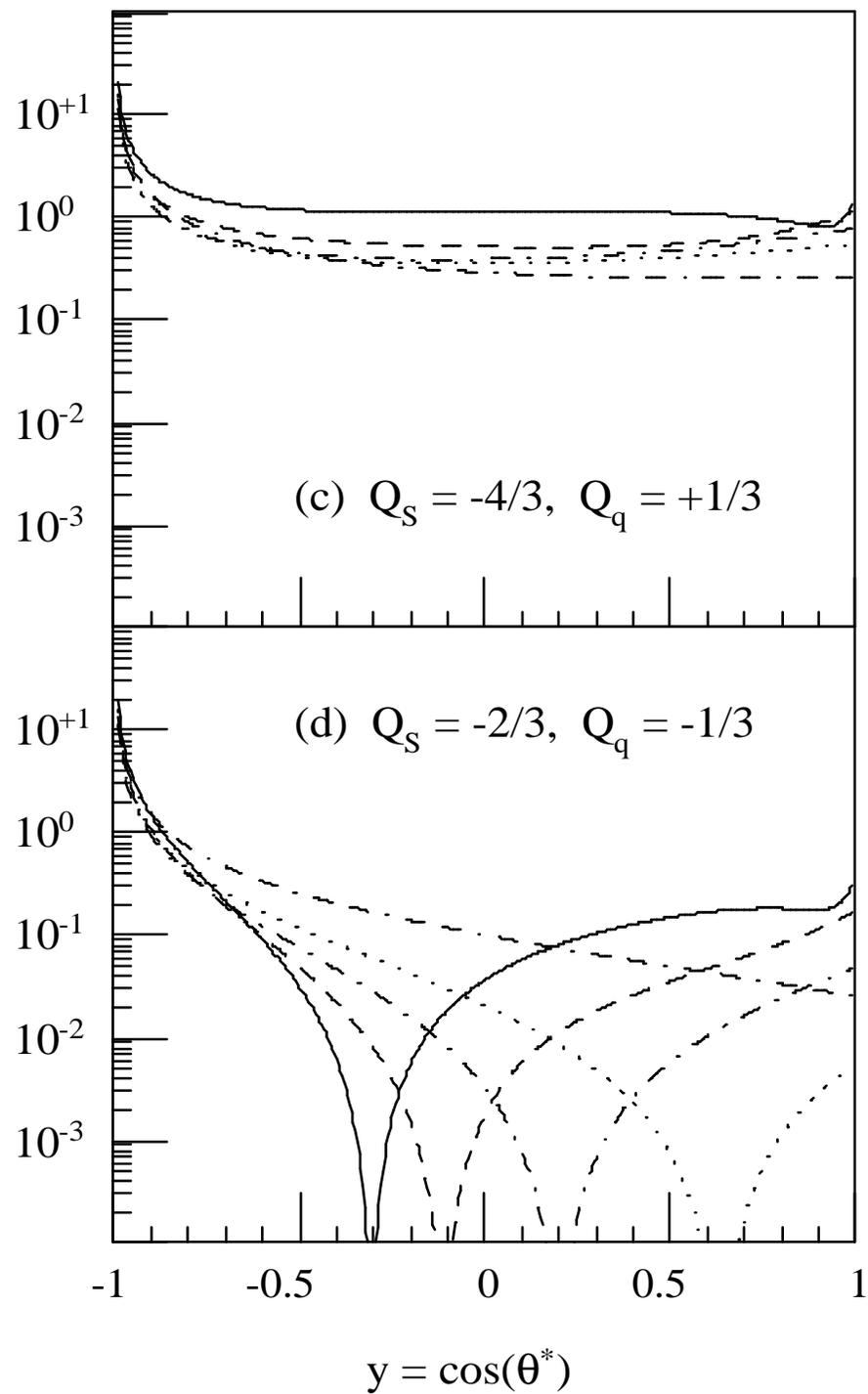
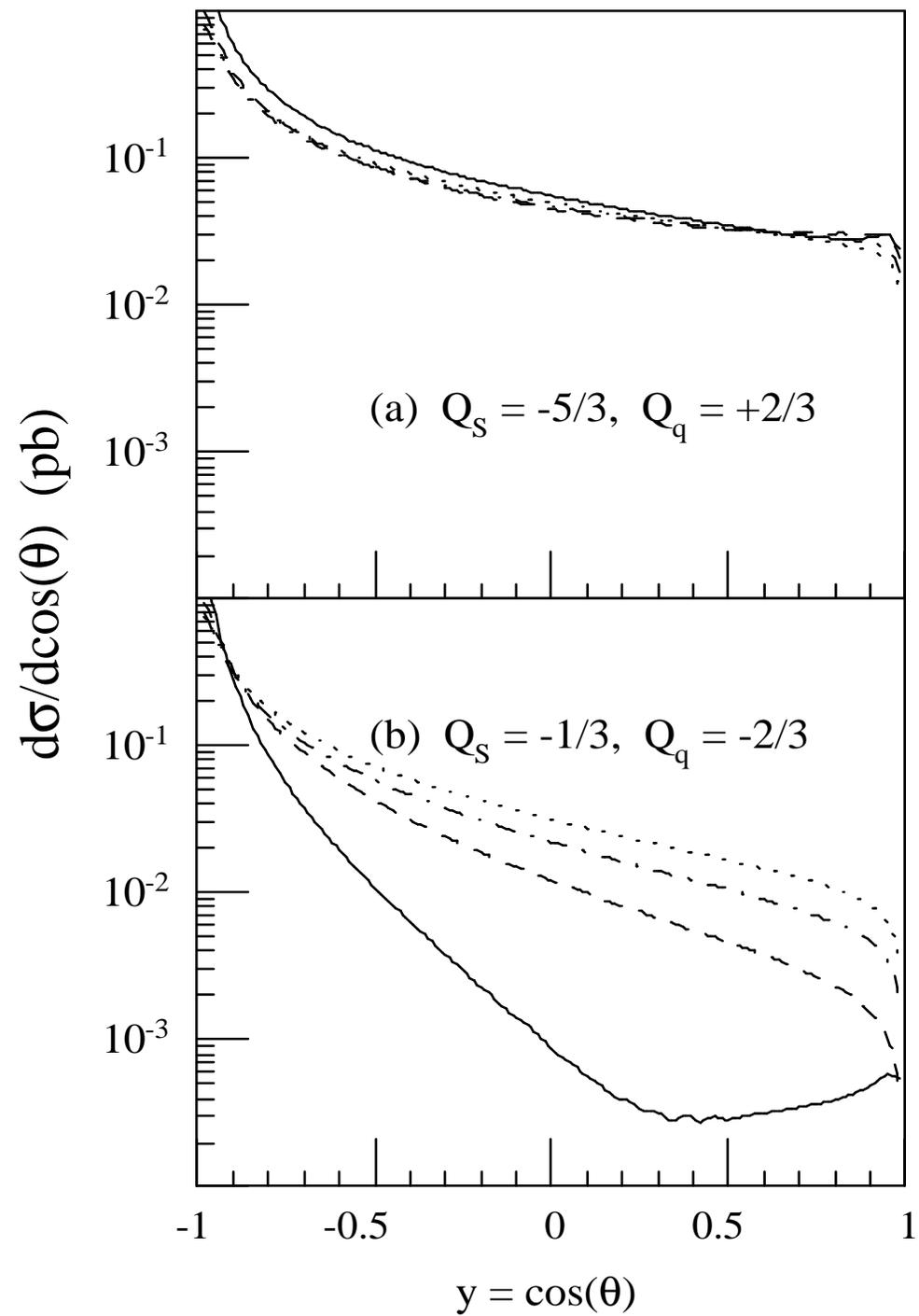
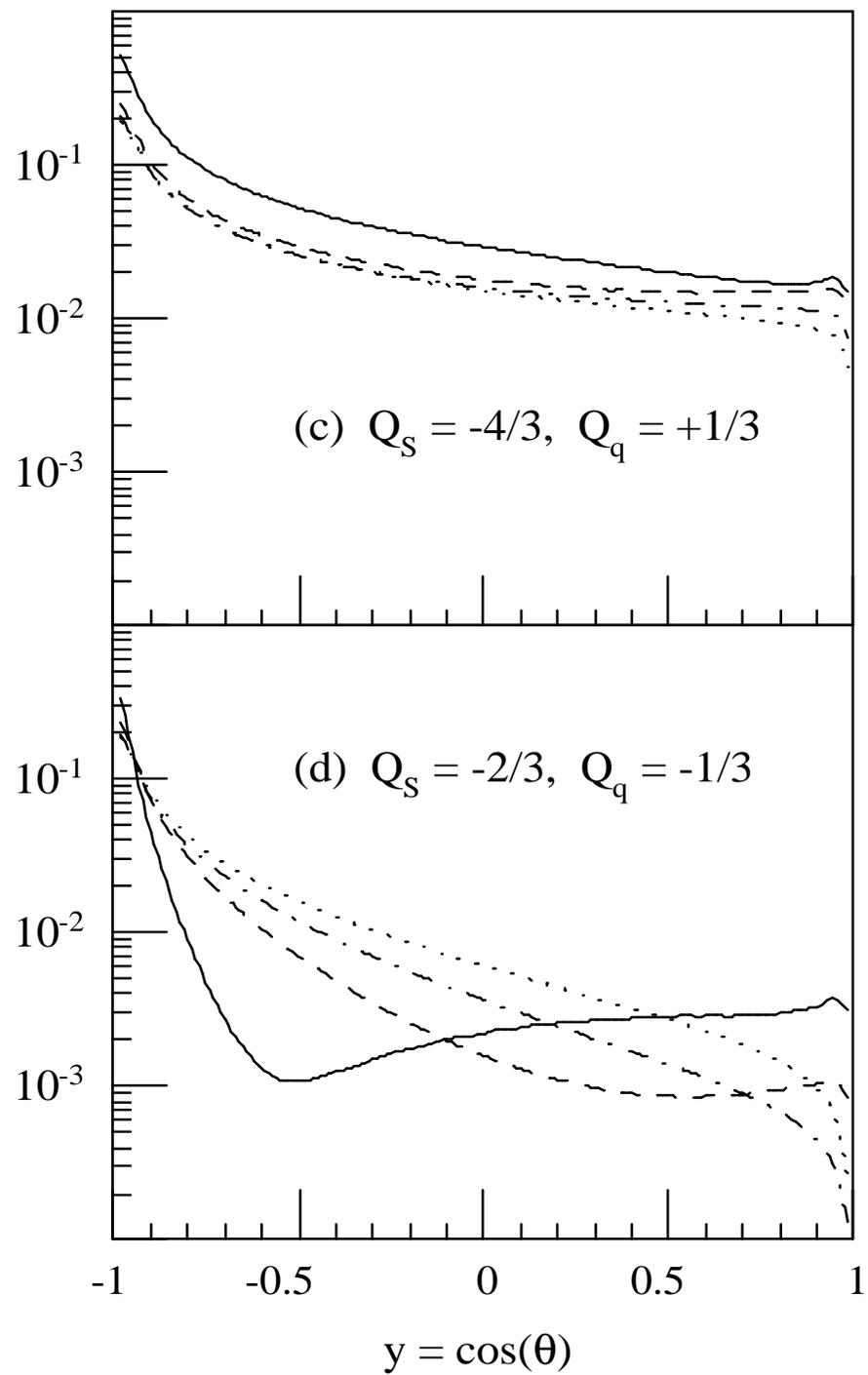
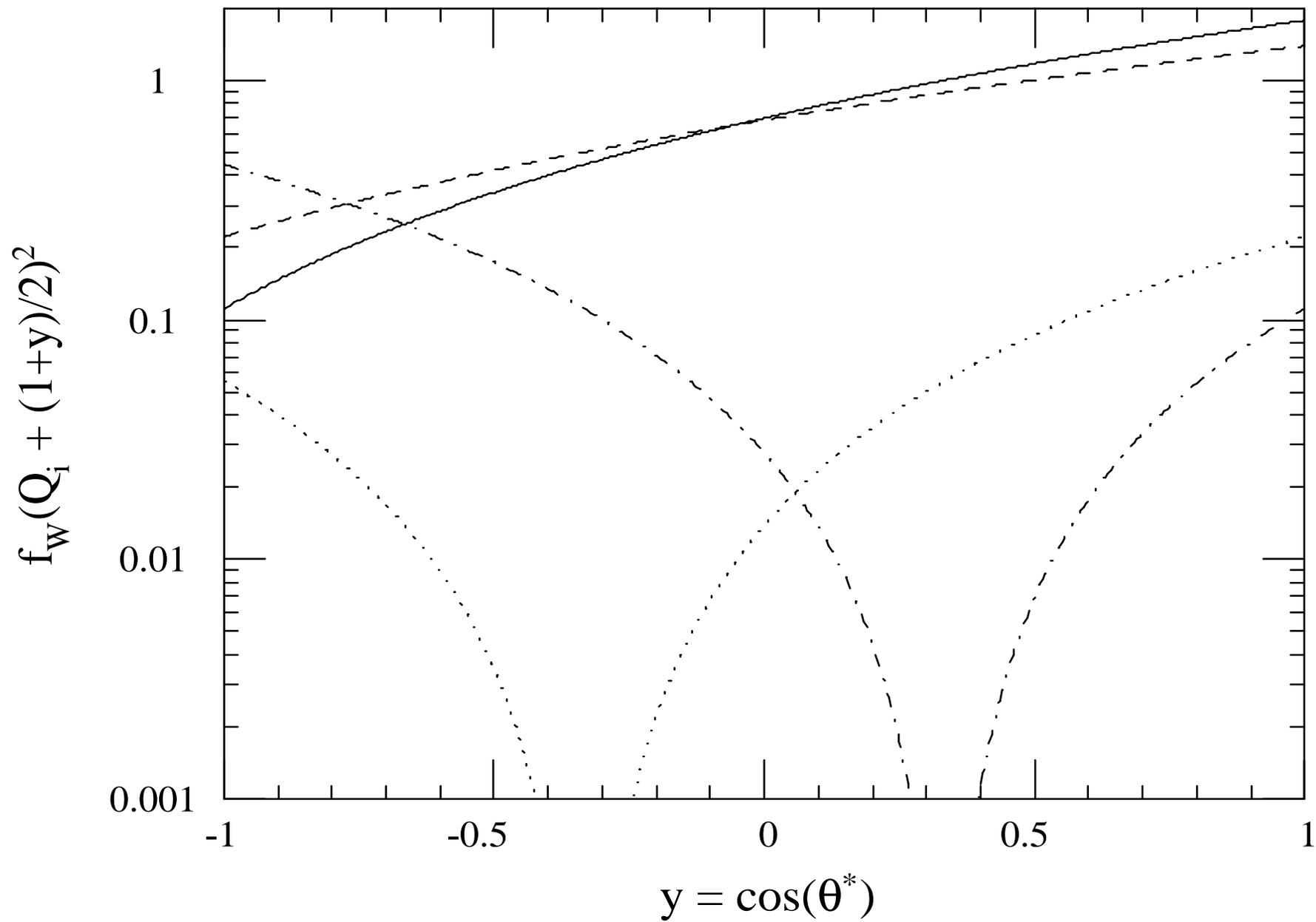
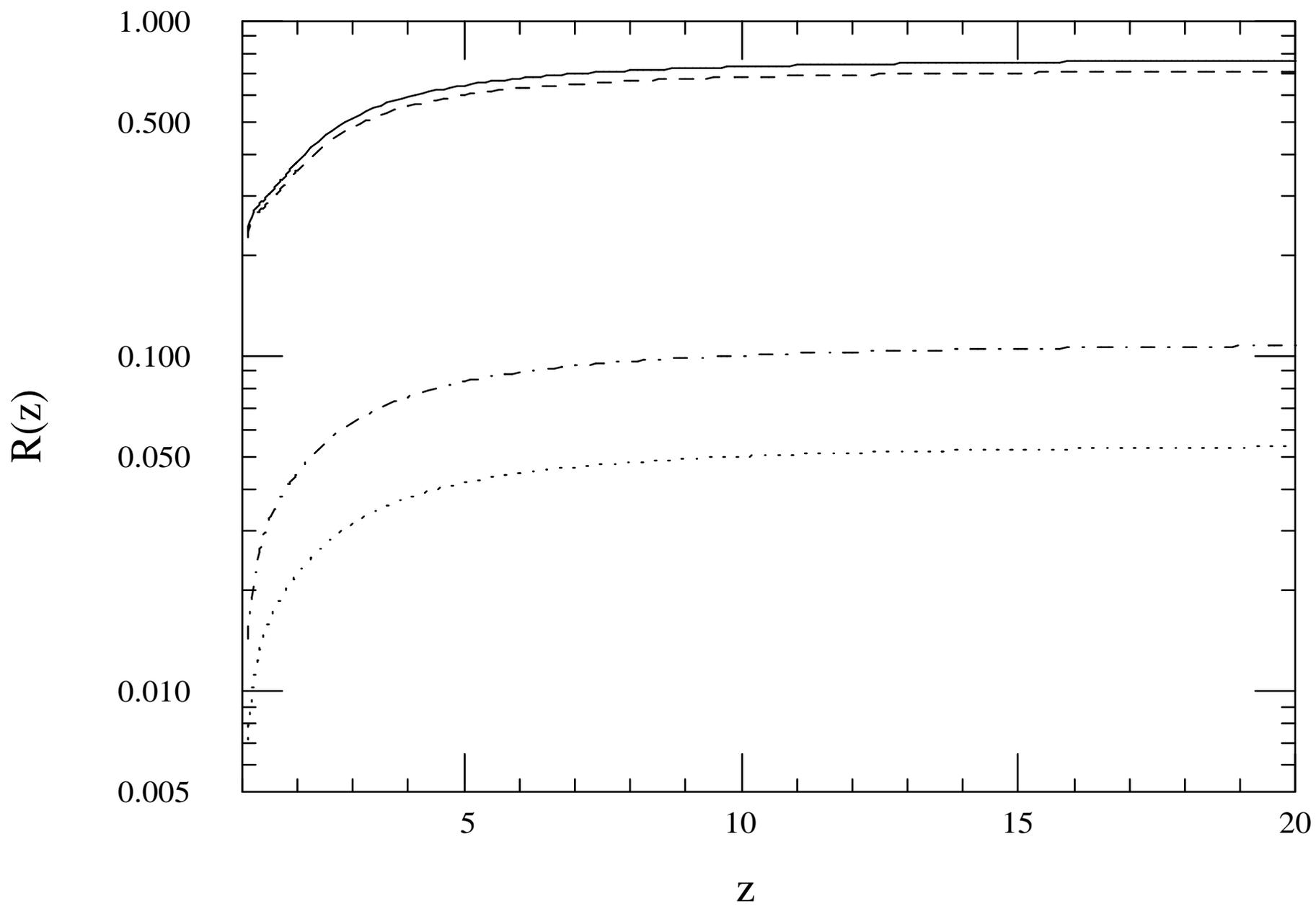
\end{document}